\pdfoutput=1
\documentclass[10pt,twocolumn]{article}
\usepackage[utf8]{inputenc}
\usepackage[margin=0.68in,top=0.62in,bottom=0.65in]{geometry}
\usepackage{amsmath,amssymb,amsfonts}
\usepackage{booktabs}
\usepackage{graphicx}
\usepackage{hyperref}
\usepackage{cite}
\usepackage{microtype}
\usepackage{xcolor}

\makeatletter
\renewcommand\section{\@startsection {section}{1}{\z@}%
  {-1.8ex \@plus -0.5ex \@minus -.2ex}%
  {0.9ex \@plus.2ex}%
  {\normalfont\large\bfseries}}
\renewcommand\subsection{\@startsection{subsection}{2}{\z@}%
  {-1.4ex\@plus -0.4ex \@minus -.2ex}%
  {0.6ex \@plus .2ex}%
  {\normalfont\normalsize\bfseries}}
\renewcommand\paragraph{\@startsection{paragraph}{4}{\z@}%
  {1.0ex \@plus0.3ex \@minus.2ex}%
  {-0.8em}%
  {\normalfont\normalsize\bfseries}}
\makeatother

\hypersetup{
  colorlinks=true,
  linkcolor=black,
  citecolor=blue,
  urlcolor=blue
}

\title{\vspace{-2.0em}\textbf{\large Ninada: Sub-Clause Split-Graph VITS Streaming and\\Capability-Routed 24~kHz Speech Synthesis Across Twelve Scripts}\vspace{-0.8em}}

\author{
  \textbf{Hemanth HM} \\
  \small\url{https://h3manth.com} \quad$\cdot$\quad \url{https://github.com/hemanth/ninada}
}

\date{\vspace{-1.5em}}

\begin{document}

\maketitle

\begin{abstract}
End-to-end VITS text-to-speech models synthesize natural speech in a single forward pass, yet standard ONNX exports execute monolithically over an entire clause, delaying Time-To-First-Audio (TTFA) to several seconds and spiking activation memory. Client-side deployment across Indic scripts faces four further bottlenecks: 138.8~MB float32 checkpoints per language strain browser caches, mobile WebGPU stalls for 6--14~s compiling convolution shaders while still running slower than real time ($1.44\times$ RTF), small WebAssembly pthread pools stall at barriers on \texttt{big.LITTLE} mobile CPUs, and mobile Web Audio pipelines drop frames or mute output under \texttt{AudioWorklet} contention and iOS \texttt{WKWebView} rules.

We present \textbf{Ninada}, an open-source 24~kHz TTS system covering English, Classical Sanskrit, and ten Indian languages (\texttt{sa}, \texttt{en}, \texttt{hi}, \texttt{kn}, \texttt{mr}, \texttt{bn}, \texttt{ta}, \texttt{te}, \texttt{ml}, \texttt{gu}, \texttt{pa}, \texttt{or}) across a capability-routed 48-checkpoint fleet: 24 studio HiFi-GAN~v1 specialists (27.9~MB \texttt{W8A32+FP16} ONNX) for desktop/WebGPU and 24 distilled \texttt{c8c8\_384} iSTFTNet specialists (21.7~MB ONNX, $2.64\times$ fewer vocoder MACs via an exact NOLA dual-window \texttt{ConvTranspose1d} inverse STFT with zero complex ops) for mobile WASM, plus a 0.0~MB zero-dependency Klatt fallback. At runtime, Ninada splits a single ONNX file in memory in $< 15$~ms into prior, flow, and windowed vocoder sub-sessions streaming 12-frame halos with raised-cosine crossfades. On $N = 480$ held-out 24~kHz studio utterances, Ninada achieves 3.22~dB 0--8~kHz MCD (3.09~dB mobile iSTFTNet; 1.33~dB 24-voice distillation MCD, vs.\ 5.17~dB for the 1,526.8~MB 16~kHz Meta MMS fleet) and 6.07~dB 8--12~kHz error, delivering 89--155~ms audible startup on desktop WebGPU and 538--649~ms zero-underrun streaming at 0.75--0.76$\times$ RTF on a physical Google Pixel 10 Pro XL.
\end{abstract}

\section{Introduction}
\label{sec:intro}

Running speech synthesis directly inside a client browser or local edge process eliminates server inference costs, network tail latency, and transcript transmission to external APIs. Screen readers, language-learning tools, bilingual voice interfaces, and offline educational readers across South Asian languages require full-band 24~kHz audio, native script fidelity across Brahmic and Latin orthographies, sub-second audible startup, and uninterrupted playback on everyday phones.

Variational Inference with adversarial learning for end-to-end Text-to-Speech (VITS)~\cite{kim2021vits} suits client deployment because it avoids autoregressive mel decoding. However, standard VITS exports hit four systems bottlenecks in web browsers:
\begin{enumerate}\setlength{\itemsep}{1pt}\setlength{\parskip}{0pt}
    \item \textbf{Utterance-Blocking Graph Execution}: Standard VITS ONNX exports chain the text encoder, stochastic duration predictor, normalizing flow, and HiFi-GAN generator~\cite{kong2020hifigan} into one static graph. No audio sample is emitted until the entire clause latent $z \in \mathbb{R}^{192 \times T}$ is upsampled by $256\times$, pushing TTFA past 2--4~s on long clauses and allocating peak activations proportional to $256 T$.
    \item \textbf{Footprint vs.\ Orthographic Fidelity}: Open checkpoints such as Meta MMS-TTS~\cite{pratap2024mms} generate 16~kHz audio (rolling off above 8~kHz) at 138.8~MB per language (1,526.8~MB across 11 Indic models). Conversely, pooling scripts into a shared Romanized model introduces cross-language accent bleed and erases script-specific contrasts such as Sanskrit vocalic sonorants, Vedic pitch marks, Malayalam atomic chillus, and Tamil/Kannada retroflex sonorants.
    \item \textbf{Mobile GPU Stalls and Compute Limits}: While desktop WebGPU runs HiFi-GAN~v1 at $\sim$0.09$\times$ Real-Time Factor (RTF), profiling pure-FP16 NHWC WebGPU on a physical Google Pixel 10 Pro XL shows 5.9--14.0~s cold shader compilation followed by $1.44\text{--}1.53\times$ warm effective RTF ($1.90\text{--}2.11\times$ NCHW), slower than real time. Meanwhile, full 28.82~GMAC/s HiFi-GAN~v1 also exceeds real time on mobile ARM WebAssembly ($1.30\times$ RTF at 6 threads).
    \item \textbf{Asymmetric CPU Barriers and Mobile Audio Quirks}: On physical 8-core \texttt{big.LITTLE} phones, small WebAssembly pthread pools stall at per-operator barriers when threads land on efficiency cores ($2\text{T} = 3.60\times$ and $3\text{T} = 2.40\times$ RTF vs.\ $1\text{T} = 1.55\times$ and $6\text{T} = 1.30\times$). Simultaneously, 2.67~ms JavaScript \texttt{AudioWorklet} callbacks contend with WASM threads on Android, while iOS \texttt{WKWebView} browsers (\texttt{CriOS}/Safari) enforce pthread memory ceilings and hardware mute-switch silencing.
\end{enumerate}

\begin{figure*}[t]
\centering
\includegraphics[width=0.86\textwidth]{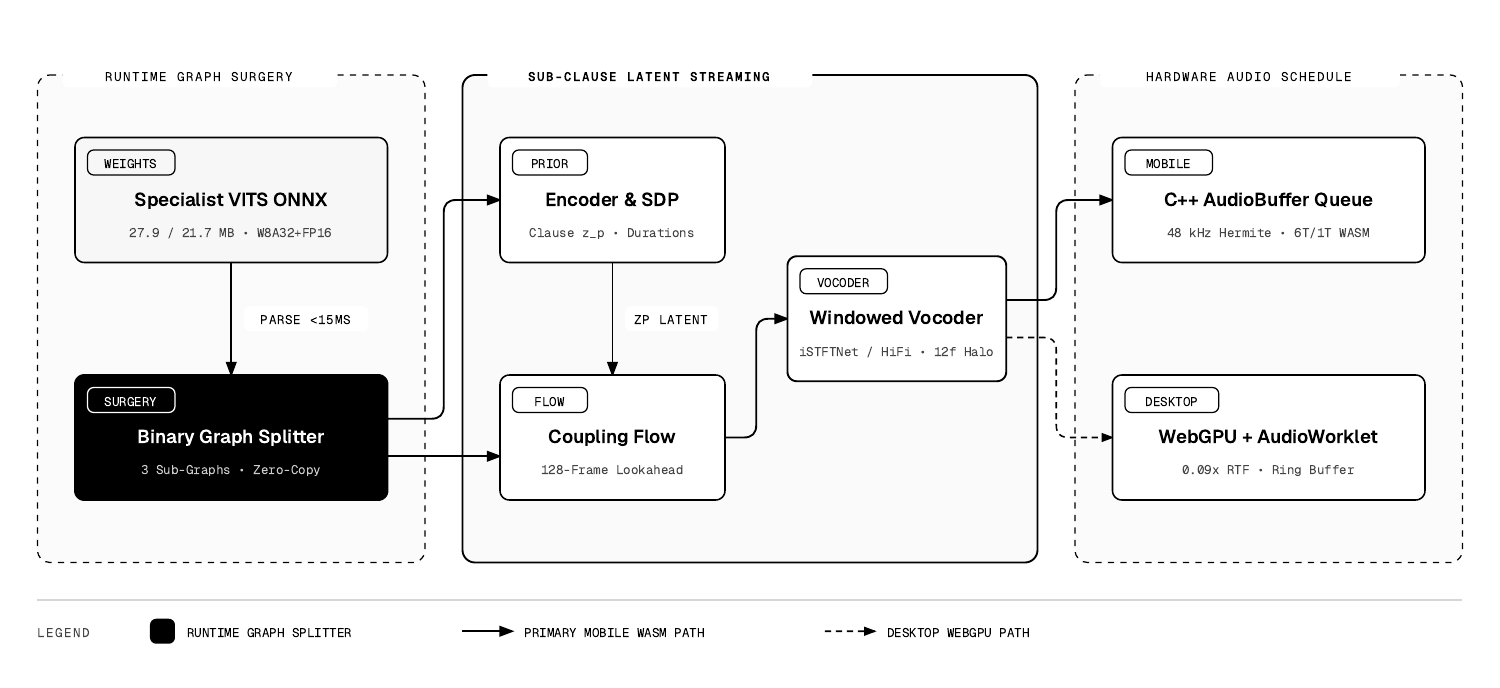}
\caption{Ninada sub-clause streaming pipeline: a single quantized VITS ONNX checkpoint (27.9~MB Studio HiFi-GAN or 21.7~MB Mobile iSTFTNet) is partitioned in memory in $< 15$~ms into prior ($G_{\text{prior}}$), coupling flow ($G_{\text{flow}}$), and windowed vocoder ($G_{\text{voc}}$) sessions. Overlapping 12-frame halos and raised-cosine crossfades feed integer-frame 48~kHz C++ \texttt{AudioBufferSourceNode} queues on mobile WASM and \texttt{AudioWorklet} ring buffers on desktop WebGPU.}
\label{fig:pipeline}
\end{figure*}

To resolve these bottlenecks without shipping multi-file voice bundles or custom ONNX Runtime builds, we present \textbf{Ninada} (Figures~\ref{fig:pipeline} and~\ref{fig:arch}). This paper makes three contributions:
\begin{itemize}\setlength{\itemsep}{1pt}\setlength{\parskip}{0pt}
    \item \textbf{Binary ONNX Graph Surgery and NOLA \texttt{ConvTranspose1d} iSTFTNet}: We introduce a zero-dependency protobuf splitter that partitions a single \texttt{W8A32+FP16} VITS checkpoint at runtime into $G_{\text{prior}}$, $G_{\text{flow}}$, and $G_{\text{voc}}$ in $< 15$~ms, alongside an ONNX-native \texttt{c8c8\_384} iSTFTNet~\cite{kaneko2022istftnet} mobile vocoder that formulates exact inverse STFT as a static real-valued \texttt{ConvTranspose1d} filterbank ($10.91$~GMAC/s, $> 138$~dB reconstruction SNR, zero complex ops) while preserving identical graph-cut signatures.
    \item \textbf{Stability Bound, \texttt{big.LITTLE} Threading, and Mobile Audio Pipeline}: We derive the minimum valid window size $V > 2 H r_0 / (1 - r_0)$ for monotonic buffer growth under halo overhead, motivating wide-pool ($6$-thread) vs.\ $1$-thread \texttt{big.LITTLE} WASM scheduling, sentence-gated clause admission when $r_0 \ge 1$, 48~kHz C1 Hermite \texttt{AudioBufferSourceNode} queuing, and synchronous iOS \texttt{AVAudioSession} playback unlock.
    \item \textbf{48-Checkpoint 12-Language Fleet and Dual-Tier Fallback}: We train 24 studio HiFi-GAN specialists (27.9~MB) and distill 24 matching mobile iSTFTNet specialists (21.7~MB, $6.4\times$ smaller than MMS-TTS) on clean \texttt{CC-BY-4.0} 24~kHz studio corpora alongside a 0.0~MB Paninian/Indic Klatt-Akshara fallback~\cite{klatt1980software}, evaluating acoustic quality ($N = 480$) and physical-device streaming telemetry.
\end{itemize}

\section{Native Orthography and Capability-Routed Architecture}
\label{sec:arch}

\begin{figure*}[t]
\centering
\includegraphics[width=0.86\textwidth]{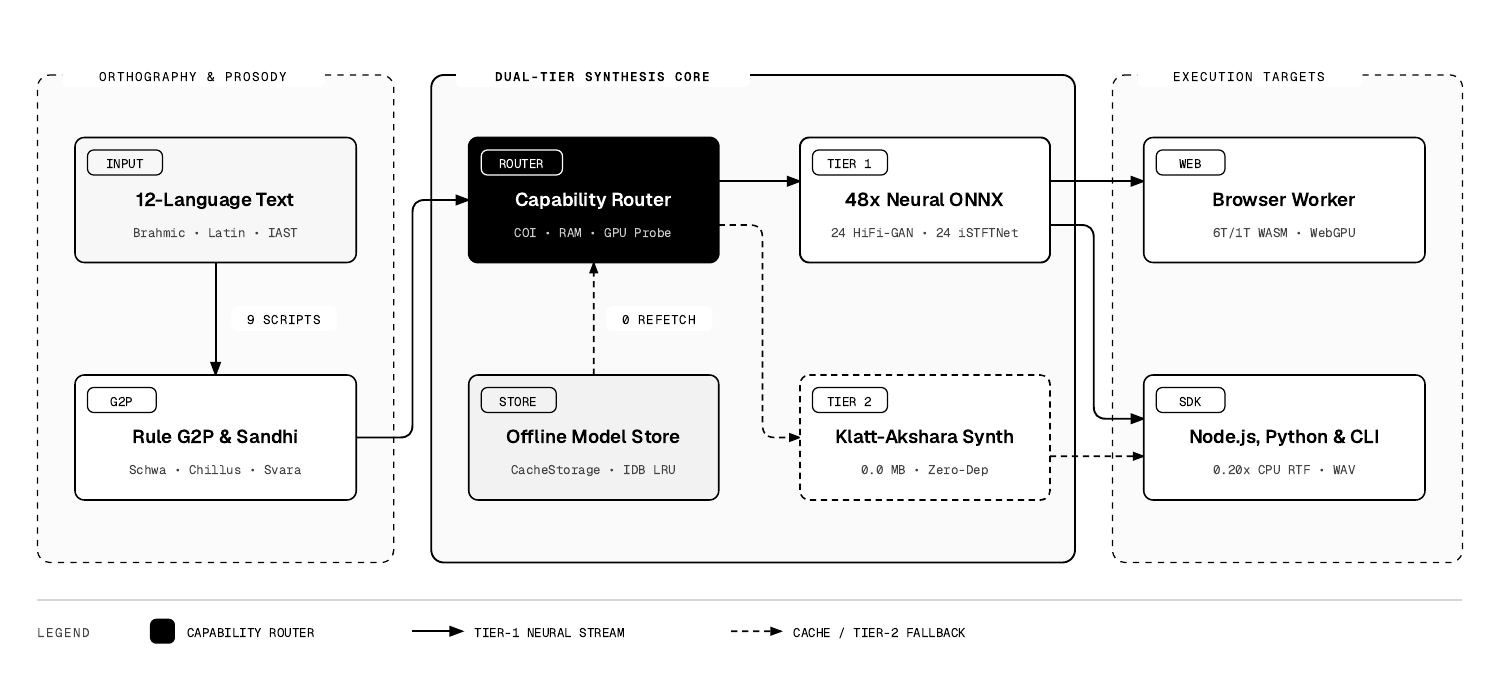}
\caption{Ninada 12-language capability-routed architecture: native Unicode and IAST inputs pass through script-specific phonological rules to a capability router backed by persistent \texttt{CacheStorage}/\texttt{IndexedDB} storage, dispatching to 27.9~MB Studio HiFi-GAN specialists on desktop/WebGPU, 21.7~MB Mobile iSTFTNet specialists on mobile WASM (Tier~1), or a 0.0~MB zero-dependency Klatt-Akshara synthesizer (Tier~2).}
\label{fig:arch}
\end{figure*}

Ninada exposes a unified streaming API across browser ES modules, Node.js, and Python (Figure~\ref{fig:arch}). Rather than routing Indic scripts through lossy Romanization, Tier-1 neural specialists tokenize native Unicode codepoints directly (\texttt{sa} 144 tokens, \texttt{en} 102, \texttt{hi} 120, \texttt{kn} 85, \texttt{mr} 144, \texttt{bn} 100, \texttt{ta} 61, \texttt{te} 110, \texttt{ml} 101, \texttt{gu} 124, \texttt{pa} 85, \texttt{or} 94), interleaving blank tokens at even indices to relax monotonic alignment constraints.

\subsection{Paninian and Pan-Indic Phonological Front-End}
For IAST Latin transliteration, prosodic inspection, and Tier-2 formant synthesis, Ninada runs a deterministic rule-based Grapheme-to-Phoneme (G2P) and syllable (\emph{akshara}) analyzer in $< 1$~ms:
\begin{itemize}\setlength{\itemsep}{1pt}\setlength{\parskip}{0pt}
    \item \textbf{Classical Sanskrit}: Preserves vocalic sonorants ($\text{\d{r}}, \bar{\text{\d{r}}}, \text{\d{l}}$), resolves homorganic nasal sandhi before stops without collapsing to generic anusv\=ara, absorbs \emph{p\=urvar\=upa} avagraha (\texttt{'}) without vowel insertion, and extracts Vedic \emph{ud\=atta}/\emph{svarita} pitch marks into prosody metadata while stripping bare combining marks from token inputs.
    \item \textbf{Modern Indo-Aryan}: Applies context-sensitive Hindi/Marathi medial \texttt{VC\_CV} and word-final schwa deletion while protecting initial syllables, conjunct clusters, and Sanskrit \emph{tatsama} suffixes~\cite{pandey2014schwa}; expands Gurmukhi \emph{tippi}/\emph{addak} gemination and localized numerals.
    \item \textbf{Dravidian Scripts}: Preserves short/long mid-vowel contrasts ($e/\bar{e}, o/\bar{o}$), retroflex approximants and laterals (Tamil \texttt{zha}, Kannada/Malayalam/Odia retroflex $l$), and normalizes Malayalam ZWJ sequences into atomic chillu codas.
\end{itemize}

\subsection{Capability-Routed Studio and Mobile Specialists}
\label{sec:specialists}
Each Tier-1 voice shares a 6-layer relative-position Transformer text encoder~\cite{shaw2018self}, a rational-quadratic neural spline Stochastic Duration Predictor (SDP)~\cite{durkan2019neural}, and a 4-layer residual affine coupling flow operating at frame hop $h = 256$ samples (10.67~ms at $f_s = 24{,}000$~Hz). The 11 Indic languages are trained on single-speaker 24~kHz anechoic studio recordings from \texttt{ai4bharat/Rasa}~\cite{varadhan2024rasa} (\texttt{CC-BY-4.0}), while English male and female specialists are trained on studio condenser recordings from \texttt{MikhailT/hifi-tts}~\cite{bakhturina2021hifitts} (\texttt{CC-BY-4.0}, resampled cleanly from 44.1~kHz to 24~kHz). Isolating one native speaker and script per checkpoint eliminates cross-language prosodic bleed.

Because desktop GPUs and mobile CPUs differ by an order of magnitude in convolution throughput, Ninada pairs each frozen prior and coupling flow with two capability-routed neural vocoders:
\begin{itemize}\setlength{\itemsep}{1pt}\setlength{\parskip}{0pt}
    \item \textbf{Studio Tier (\texttt{onnx/*.onnx}, 27.9~MB)}: Desktop WebGPU, desktop multi-core WASM, Node.js, and Python load a full time-domain HiFi-GAN~v1 decoder~\cite{kong2020hifigan} with upsampling rates $[8, 8, 2, 2]$ and initial width $512$ ($28.82$~GMAC/s per synthesized second).
    \item \textbf{Mobile Tier (\texttt{onnx\_mobile/*.onnx}, 21.7~MB)}: Mobile browsers and low-RAM hosts ($\le 4$~GB) load a distilled \texttt{c8c8\_384} iSTFTNet decoder~\cite{kaneko2022istftnet} that retains only the first two neural upsampling stages $[8, 8]$ at width $384$ ($64\times$ neural upsampling to 6~kHz frame rate) and delegates the final $4\times$ upsampling to an inverse STFT head ($N_{\text{fft}} = 16, h_{\text{stft}} = 4$, $K = 9$ frequency bins), cutting vocoder compute by $2.64\times$ to $10.91$~GMAC/s.
\end{itemize}

Standard ONNX \texttt{InverseSTFT} and complex-tensor operators lack portable WebAssembly SIMD kernels in \texttt{onnxruntime-web}. Observing that a periodic Hann analysis window $w[n]$ with $N_{\text{fft}} = 4 h_{\text{stft}}$ satisfies the constant-overlap-add identity $\sum_{m} w^2[n - m h_{\text{stft}}] \equiv \frac{3}{2}$ for all $n$, we define the exact Nonzero-Overlap-Add (NOLA) dual window $\tilde{w}[n] = \frac{2}{3} w[n]$ and express inverse STFT synthesis from predicted log-magnitude $\hat{M} \in \mathbb{R}^{K \times 64T}$ and phase $\hat{\Phi} \in \mathbb{R}^{K \times 64T}$ as a single static real-valued transposed convolution:
\begin{equation}
\hat{y} = \operatorname{ConvT1d}_{h_{\text{stft}}}\!\left(\begin{bmatrix}e^{\hat{M}} \cos \hat{\Phi} \\ e^{\hat{M}} \sin \hat{\Phi}\end{bmatrix},\, \mathbf{B}\right)
\label{eq:istft}
\end{equation}
where $\operatorname{ConvT1d}_{h_{\text{stft}}}$ has stride $h_{\text{stft}} = 4$ and padding $(N_{\text{fft}} - h_{\text{stft}})/2 = 6$, and $\mathbf{B} \in \mathbb{R}^{2K \times 1 \times N_{\text{fft}}}$ stacks fixed cosine and sine Fourier basis filters scaled by $\tilde{w}[n]$ ($1/N_{\text{fft}}$ at DC/Nyquist and $2/N_{\text{fft}}$ on interior bins). Equation~\ref{eq:istft} achieves $> 138$~dB reconstruction SNR against \texttt{torch.istft} with zero runtime divisions, complex types, or custom operators. Each mobile vocoder is distilled for 30,000 steps on paired teacher latents $(z, y_{\text{teacher}})$ and studio posterior latents $(z_{\text{post}}, y_{\text{studio}})$ using multi-resolution STFT, sub-band magnitude, mel-$L_1$, and Multi-Period Discriminator feature-matching losses, then surgically grafted into the frozen specialist graph while preserving the entry \texttt{conv\_pre} tensor name.

Mixed \texttt{W8A32+FP16} compression quantizes encoder and flow projections to dynamic INT8 (FP32 accumulation) while storing vocoder convolutions and Fourier basis weights in IEEE FP16, shrinking Studio checkpoints from 112.0~MB to 27.92~MB and Mobile checkpoints to 21.68~MB ($6.4\times$ smaller than MMS-TTS). Specialists persist in \texttt{CacheStorage} with an \texttt{IndexedDB} LRU fallback ($0$ network refetches on reload), and automatically fall back offline to Tier~2: a 0.0~MB cascade Klatt synthesizer~\cite{klatt1980software} with $-58$~dBFS inter-clause room tone.

\section{Sub-Clause Split-Graph Streaming}
\label{sec:streaming}

\subsection{Runtime Binary ONNX Graph Partitioning}
Shipping three separate ONNX files across 48 neural checkpoints would triple asset management and cache transactions. Instead, Ninada ships one self-contained \texttt{.onnx} file per voice tier and performs binary protobuf surgery in JavaScript or Python during session creation (Figure~\ref{fig:pipeline}).

Given a serialized \texttt{ModelProto} buffer, the splitter scans top-level \texttt{GraphProto} wire tags without external protobuf libraries, indexing \texttt{NodeProto} producers, \texttt{Initializer} byte spans, and \texttt{ValueInfoProto} descriptors. It locates two topological cut boundaries shared identically by both HiFi-GAN and iSTFTNet checkpoints:
\begin{enumerate}\setlength{\itemsep}{1pt}\setlength{\parskip}{0pt}
    \item Prior latent tensor $z_p \in \mathbb{R}^{1 \times 192 \times T}$ and mask $m_y \in \mathbb{R}^{1 \times 1 \times T}$ emitted after SDP duration expansion;
    \item Inverted flow latent tensor $z \in \mathbb{R}^{1 \times 192 \times T}$ feeding the entry $1 \times 1$ \texttt{conv\_pre} layer of the waveform decoder.
\end{enumerate}
Backward reachability partitions nodes and raw initializer slices into three valid \texttt{ModelProto} buffers ($G_{\text{prior}}, G_{\text{flow}}, G_{\text{voc}}$) in $< 15$~ms and immediately releases the monolithic buffer so worker RAM remains bounded to a single model footprint.

\subsection{Receptive-Field Halos and Seam Continuity}
Slicing $z$ into disjoint frame blocks causes boundary clicks because dilated residual stacks and transposed convolutions span a finite temporal receptive field. For window $k$ covering valid frames $[f_k, f_{k+1})$ of length $V_k = f_{k+1} - f_k$, Ninada evaluates an extended latent slice $[f_k - L_k, f_{k+1} + R_k)$ padded with symmetric context halos:
\begin{equation}
L_k = \min(f_k, H), \qquad R_k = \min(T - f_{k+1}, H)
\label{eq:halo}
\end{equation}
with $H = 12$ latent frames ($3{,}072$ samples, or 128~ms per side), while $G_{\text{flow}}$ runs in bounded blocks with a $H_{\text{flow}} = 24$-frame halo. After $G_{\text{voc}}$ emits $(V_k + L_k + R_k) h$ samples, we trim the $L_k h$ left-halo samples, retain $N_{\text{xf}} = 512$ samples (21.3~ms) from the right halo as overlap tail $\tau^{(k)}$, and blend window $k+1$ via a C1 raised-cosine crossfade:
\begin{equation}
y^{(k+1)}[n] = (1 - \alpha_n)\,\tau^{(k)}[n] + \alpha_n\,\hat{y}^{(k+1)}[L_{k+1} h + n]
\label{eq:xfade}
\end{equation}
for $0 \le n < N_{\text{xf}}$, where $\alpha_n = \frac{1}{2}\!\left(1 - \cos\!\frac{\pi (n + 0.5)}{N_{\text{xf}}}\right)$, followed by a first-order DC blocker ($R = 0.995$).

\subsection{Stability Bound, \texttt{big.LITTLE} Scaling, and Mobile Audio}
\label{sec:bound}
On desktop WebGPU, $G_{\text{voc}}$ uses two static frame buckets ($44$ frames on Chunk~0 and $112$ steady-state) so WGSL shaders compile only twice at $\sim$0.09$\times$ RTF. On mobile browsers, WebGPU is bypassed because shader compilation stalls startup for 5.9--14.0~s while warm 1D transposed convolutions still run at $1.44\text{--}1.53\times$ RTF on phone GPUs; ARM Neon WASM SIMD compiles in $< 90$~ms and executes exact unpadded window lengths $W_k = V_k + L_k + R_k$.

Let $r_0$ denote the raw per-frame compute ratio on WASM (wall time to flow-invert and vocode one latent frame divided by 10.67~ms). An interior window with $V$ valid frames evaluates $V + 2H$ frames ($2H = 24$). Playback cushion grows monotonically across interior windows ($\Delta B > 0$) if and only if:
\begin{equation}
(V + 2H) r_0 < V \iff r_0 < 1 \;\;\text{and}\;\; V > \frac{2 H r_0}{1 - r_0}
\label{eq:bound}
\end{equation}
Equation~\ref{eq:bound} governs both model architecture and scheduler design. First, when $r_0 \ge 1$---as occurs with full HiFi-GAN~v1 on physical phone CPUs ($r_0 \approx 1.08$, effective RTF $1.30\times$)---no finite window $V$ can prevent mid-clause buffer depletion; distilling $G_{\text{voc}}$ to \texttt{c8c8\_384} iSTFTNet lowers raw mobile WASM compute to $r_0 \approx 0.56$, restoring a valid steady-state regime ($V > 30.5$ frames). Second, even when $r_0 < 1$, overly narrow windows ($V = 20 \Rightarrow 55\%$ halo overhead) invert effective RTF above $1.0\times$, whereas overly wide initial windows inflate TTFA.

Ninada combines four mechanisms for cross-browser mobile streaming:
\begin{enumerate}\setlength{\itemsep}{1pt}\setlength{\parskip}{0pt}
    \item \textbf{Pre-Computed Flow + Window Ramp}: Clause~0 vocodes $W_0 = 68$ frames ($V_0 = 56$, 597~ms audio) while pre-computing $128$ frames of $G_{\text{flow}}$. Chunk~1 ($V_1 = 60, W_1 = 84$) executes zero flow ops, followed by $V_2 = 80$ and steady-state $V_{\ge 3} = 96$ ($W = 120$, $20\%$ halo overhead).
    \item \textbf{Bimodal \texttt{big.LITTLE} Thread Policy}: Homogeneous desktop CPU throttling masks mobile barrier contention. On a physical 8-core Pixel 10 Pro XL, 2--3 WASM threads run $1.5\text{--}2.3\times$ \emph{slower} than 1 thread because Emscripten pthreads block on efficiency cores at every convolution barrier, whereas a 6-thread pool saturates prime/mid cores. Under Cross-Origin Isolation, Ninada therefore assigns $\min(6, C - 2)$ threads when logical cores $C \ge 8$ and $1$ thread when $C < 8$; on iOS/iPadOS \texttt{WKWebView}, it skips COI Service Workers and pins $1$ thread to avoid pthread Jetsam termination.
    \item \textbf{Sentence-Gated Clause Admission}: On single-threaded or thermally throttled devices where observed runtime RTF $\hat{r} > 1$, Ninada admits subsequent clauses into hardware playback once queued lead $B_{\text{ahead}}$ plus synthesized clause audio $B_{\text{queued}}$ satisfy $B_{\text{ahead}} + B_{\text{queued}} \ge D_{\text{rem}} \max(0, \hat{r} - 1) + W_{\text{next}}$, shifting any unavoidable compute wait cleanly onto natural inter-sentence pauses with zero mid-word underruns.
    \item \textbf{Hardware-Rate C++ Queue and iOS Session Unlock}: Mobile browsers upsample each 24~kHz chunk on arrival ($< 0.2$~ms) to hardware rate (48~kHz) via a stateful 4-point C1 Hermite interpolator and schedule integer-frame \texttt{AudioBufferSourceNode} buffers on Blink/WebKit's native C++ audio thread. On iOS Safari and Chrome (\texttt{CriOS}), Ninada unlocks \texttt{AudioContext} synchronously inside \texttt{touchend}/\texttt{click} via a 1-frame buffer and activates \texttt{navigator.audioSession.type = 'playback'} alongside a silent WAV bridge so speech plays even when the hardware Ring/Silent switch is muted.
\end{enumerate}

\section{Evaluation}
\label{sec:eval}

\subsection{12-Language Acoustic Quality \texorpdfstring{($N = 480$)}{(N = 480)}}
We evaluate Ninada on $N = 480$ held-out 24,000~Hz single-speaker studio utterances ($40$ per language across all 12 languages), strictly disjoint from the $27{,}600$ training and $480$ validation utterances. Following Kubichek~\cite{kubichek1993mcd}, Dynamic Time Warping (DTW) Mel-Cepstral Distance (MCD) is computed over the shared 0--8~kHz band for fair comparison against 16~kHz baselines, while 8--12~kHz spectral error measures upper-band sibilance reconstruction. Confidence intervals use $B = 1{,}000$ non-parametric bootstrap resamples.

\begin{table*}[t]
\centering
\caption{12-Language Held-Out Studio Benchmark ($N = 480$ utterances, $40$ per language; $B = 1{,}000$ bootstrap 95\% CIs in brackets). $^\dagger$Mobile \texttt{c8c8\_384} iSTFTNet exhibits 1.33~dB mean teacher-student MCD across all 24 male/female voices ($100.0\%$ voiced) and 474.4~ms first-chunk TTFA on a physical Google Pixel 10 Pro XL.}
\label{tab:leaderboard}
\small
\resizebox{\textwidth}{!}{%
\begin{tabular}{@{}lccccccc@{}}
\toprule
\textbf{System} & \textbf{Rate} & \textbf{Active Size ($\downarrow$)} & \textbf{Coverage ($\uparrow$)} & \textbf{TTFA p50 ($\downarrow$)} & \textbf{MCD 0--8~kHz (dB, $\downarrow$)} & \textbf{8--12~kHz Err (dB, $\downarrow$)} & \textbf{Dur MAE ($\downarrow$)} \\
\midrule
Base \texttt{mms-tts-hin} (Single VITS) & 16~kHz & 138.8~MB & 29.4\% & 0.9~ms$^*$ & 7.98 [7.81, 8.16] & 65.16~dB & 3.49~s \\
Meta MMS 11-Specialist Fleet & 16~kHz & 1526.8~MB & 100.0\% & 201.1~ms & 5.17 [5.12, 5.22] & 64.97~dB & 1.04~s \\
Ninada Warm-Start (\texttt{step=0}) & 24~kHz & 112.0~MB & 100.0\% & 191.0~ms & 4.58 [4.54, 4.62] & 10.93~dB & 0.93~s \\
Ninada Tier-2 (Zero-Dep Klatt) & 24~kHz & \textbf{0.0~MB} & \textbf{100.0\%} & 110.2~ms & 10.53 [10.43, 10.64] & 42.38~dB & 0.88~s \\
\textbf{Ninada Studio FP32 (HiFi-GAN)} & \textbf{24~kHz} & 112.0~MB & \textbf{100.0\%} & \textbf{43.3~ms} & 3.22 [3.19, 3.25] & 6.08~dB & \textbf{0.58~s} \\
\textbf{Ninada Studio ONNX (\texttt{W8A32+FP16})} & \textbf{24~kHz} & 27.9~MB & \textbf{100.0\%} & 373.0~ms & 3.22 [3.19, 3.25] & \textbf{6.07~dB} & \textbf{0.58~s} \\
\textbf{Ninada Mobile ONNX (\texttt{c8c8\_384} iSTFTNet)}$^\dagger$ & \textbf{24~kHz} & \textbf{21.7~MB} & \textbf{100.0\%} & 474.4~ms & \textbf{3.09} (1.33 distill) & 6.28~dB & \textbf{0.58~s} \\
\bottomrule
\end{tabular}%
}
\end{table*}

Table~\ref{tab:leaderboard} summarizes overall acoustic and latency metrics ($^*$single-checkpoint \texttt{mms-tts-hin} fails on non-Devanagari scripts, voicing 29.4\% of utterances). Relative to the 1,526.8~MB Meta MMS 11-checkpoint fleet, Ninada's 27.9~MB Studio ONNX specialists lower mean 0--8~kHz MCD by 1.95~dB ($5.17 \to 3.22$~dB), cut 8--12~kHz spectral error from 64.97~dB to 6.07~dB via full-band 24~kHz synthesis, and reduce duration MAE from 1.04~s to 0.58~s, matching FP32 within 0.01~dB ($0.58$~dB direct waveform parity MCD, $r = 0.9999$ duration correlation, $0.20\times$ desktop CPU RTF).

Distilling the waveform decoder into the 21.7~MB \texttt{c8c8\_384} iSTFTNet mobile architecture preserves identical token durations ($\text{MAE} = 0.58$~s) while achieving 1.33~dB mean teacher-student MCD across all 24 male and female voices (1.27~dB male, 1.39~dB female, 1.26~dB 8--12~kHz deviation vs.\ teacher, $100.0\%$ voiced; Table~\ref{tab:perlang}). Because explicit sub-band magnitude supervision smooths high-variance adversarial phase jitter on unvoiced transitions, the mobile student attains 3.09~dB mean 0--8~kHz DTW MCD against studio ground truth (6.28~dB 8--12~kHz error) while using $62.2\%$ fewer vocoder MACs.

\begin{table}[ht]
\centering
\caption{Per-Language 0--8~kHz Ground-Truth MCD and 24-Voice iSTFTNet Teacher-Student Distillation MCD (dB, $\downarrow$, $N = 40$ per voice).}
\label{tab:perlang}
\small
\resizebox{\columnwidth}{!}{
\begin{tabular}{@{}lccccc@{}}
\toprule
\textbf{Language} & \textbf{MMS} & \textbf{Warm} & \textbf{Studio} & \textbf{Mobile} & \textbf{Distill (M / F)} \\
\midrule
Sanskrit (\texttt{sa}) & 5.32 & 4.79 & 3.47 & \textbf{3.17} & 1.41 / 1.38 \\
Hindi (\texttt{hi}) & 4.85 & 4.33 & 3.33 & \textbf{3.07} & 1.31 / 1.18 \\
Kannada (\texttt{kn}) & 5.04 & 4.17 & \textbf{3.30} & \textbf{3.30} & 1.40 / 1.33 \\
Tamil (\texttt{ta}) & 4.87 & 4.69 & 3.16 & \textbf{2.76} & 1.47 / 1.12 \\
Telugu (\texttt{te}) & 5.57 & 4.94 & 3.56 & \textbf{3.08} & 1.45 / 1.49 \\
Malayalam (\texttt{ml}) & 4.88 & 4.68 & 3.29 & \textbf{2.83} & 1.37 / 1.34 \\
English (\texttt{en}) & 5.94 & 4.34 & \textbf{3.07} & 3.69 & 1.20 / 1.19 \\
Marathi (\texttt{mr}) & 4.62 & 4.55 & 3.06 & \textbf{2.95} & 1.15 / 1.31 \\
Bengali (\texttt{bn}) & 4.68 & 4.08 & 3.22 & \textbf{3.15} & 1.08 / 1.77 \\
Gujarati (\texttt{gu}) & 5.56 & 4.57 & 3.25 & \textbf{3.19} & 1.18 / 1.23 \\
Punjabi (\texttt{pa}) & 5.23 & 4.76 & 2.88 & \textbf{2.85} & 1.00 / 1.21 \\
Odia (\texttt{or}) & 5.49 & 5.07 & 3.05 & \textbf{2.99} & 1.26 / 2.09 \\
\midrule
\textbf{12-Lang Mean} & 5.17 & 4.58 & 3.22 & \textbf{3.09} & \textbf{1.27 / 1.39} \\
\bottomrule
\end{tabular}
}
\end{table}

\subsection{Split-Graph Seam Continuity and Real-Device Streaming}
Across an adversarial seam audit spanning all 12 languages ($141$ streamed chunks on punctuated passages and $376$ chunks on 350--393 character unpunctuated clauses), maximum boundary sample step $|\Delta y|$ remains below $3.29 \times 10^{-5}$ with DC offset $< 8.6 \times 10^{-7}$ and $> 38$~dB split-vs-monolithic SNR parity.

\begin{table}[ht]
\centering
\caption{End-to-End Browser Streaming Telemetry Across Desktop, Physical Phone Silicon (Google Pixel 10 Pro XL), and Throttled Profiles.}
\label{tab:browser}
\small
\resizebox{\columnwidth}{!}{
\begin{tabular}{@{}lccccc@{}}
\toprule
\textbf{Device / Profile} & \textbf{Tier \& Backend} & \textbf{Audible} & \textbf{RTF} & \textbf{Underruns} & \textbf{Min Cushion} \\
\midrule
Desktop Chrome (EN 15.6s) & Studio WebGPU & 147~ms & 0.092 & \textbf{0} & 240~ms \\
Desktop Chrome (KN 14.4s) & Studio WebGPU & 89~ms & 0.093 & \textbf{0} & 218~ms \\
Pixel 10 Pro XL (HiFi-GAN) & Studio GPU FP16 & 7727~ms & 1.477 & $>$3 & 0~ms \\
Pixel 10 Pro XL (HiFi-GAN) & Studio WASM-6T & 1085~ms & 1.301 & $>$2 & 0~ms \\
\textbf{Pixel 10 Pro XL (EN 7.2s)} & \textbf{Mobile WASM-6T} & \textbf{649~ms} & \textbf{0.749} & \textbf{0} & \textbf{562~ms} \\
\textbf{Pixel 10 Pro XL (EN 9.9s)} & \textbf{Mobile WASM-6T} & \textbf{538~ms} & \textbf{0.761} & \textbf{0} & \textbf{221~ms} \\
Android 2$\times$ Sim (SA 12.7s) & Studio WASM & 423~ms & 0.551 & \textbf{0} & 285~ms \\
Android 2$\times$ Sim (KN 14.4s) & Studio WASM & 394~ms & 0.629 & \textbf{0} & 263~ms \\
\bottomrule
\end{tabular}
}
\end{table}

Table~\ref{tab:browser} contrasts desktop WebGPU, simulated CPU throttling, and physical smartphone hardware (Google Pixel 10 Pro XL over Wireless ADB + Chrome DevTools). While $2\times$ desktop-throttled Playwright runs Studio HiFi-GAN at $0.55\text{--}0.63\times$ RTF, physical smartphone silicon exposes why capability-routed distillation is essential: Studio HiFi-GAN runs at $1.48\times$ RTF on mobile FP16 WebGPU (after a 7.7~s shader stall) and $1.30\times$ RTF on 6-thread mobile WASM, violating Equation~\ref{eq:bound}. Routing the Pixel 10 Pro XL to the 21.7~MB \texttt{c8c8\_384} iSTFTNet specialist on 6-thread WASM SIMD cuts end-to-end streaming RTF to $0.75\text{--}0.76\times$ ($362\text{--}366$~ms Chunk~0 vocode), starting audible speech in 538--649~ms with 0 underruns, 0 late buffer starts, and 221--562~ms minimum hardware lead across multi-sentence passages, while keeping main-thread JS heap within 3.5--11.0~MB across six consecutive language switches.

\section{Discussion \& Limitations}
\label{sec:discussion}

\paragraph{Single-Threaded and Un-Isolated Mobile Browsers.}
Multi-threaded WebAssembly requires \texttt{SharedArrayBuffer} via Cross-Origin Isolation (\texttt{COOP}/\texttt{COEP} headers or a Service Worker). On iOS/iPadOS \texttt{WKWebView} browsers (Safari and Chrome \texttt{CriOS}) where Service Worker pthread pools risk Jetsam memory kills, and in embedded social-app WebViews blocking Service Workers, ONNX Runtime executes on 1 WASM thread. Even with \texttt{c8c8\_384} iSTFTNet lowering 1-thread mobile RTF near $0.95\text{--}1.15\times$, thermal throttling can push $r_0 \ge 1$; Ninada's sentence-gated clause admission prevents mid-word clicks or dropouts by absorbing any deficit cleanly into natural inter-sentence pauses.

\paragraph{Single-Style Studio Voices and Asset Footprint.}
Each specialist models one anechoic male or female studio voice per language without zero-shot speaker cloning, trading multi-speaker conditioning for compact 21.7--27.9~MB client caching across 24 voices (48 capability-routed files on Hugging Face, with zero simultaneous download of both tiers on a single device).

\section{Related Work}
\label{sec:related}

\paragraph{End-to-End Indic TTS and Fast Neural Vocoders.}
VITS~\cite{kim2021vits} unifies variational alignment, normalizing flows~\cite{durkan2019neural}, and HiFi-GAN~\cite{kong2020hifigan}. Meta MMS~\cite{pratap2024mms} scaled 16~kHz VITS to 1,100+ languages at 138.8~MB per checkpoint, while AI4Bharat Rasa~\cite{varadhan2024rasa} established studio-grade 24~kHz Indic corpora. Kaneko et al.~\cite{kaneko2022istftnet} introduced iSTFTNet to replace outer HiFi-GAN upsamplers with inverse STFT. Ninada adapts iSTFTNet inside VITS via an exact NOLA dual-window \texttt{ConvTranspose1d} synthesis filterbank compatible with standard WebAssembly SIMD and surgical runtime ONNX graph splitting.

\paragraph{Client-Side Speech Synthesis.}
Formant synthesizers derived from Klatt~\cite{klatt1980software} and eSpeak run in kilobytes with millisecond latency but high spectral distortion ($> 10$~dB MCD). Browser neural ports such as Piper and Sherpa-ONNX synthesize full clauses monolithically per worker call. Ninada unifies sub-clause split-graph neural streaming with capability-routed desktop/mobile vocoders and a zero-dependency Klatt-Akshara fallback.

\section{Conclusion}
\label{sec:conclusion}

We presented \textbf{Ninada}, a 24,000~Hz streaming text-to-speech system spanning 12 Indic and English scripts across browser, Node.js, and Python runtimes. By combining $< 15$~ms binary ONNX graph surgery, capability-routed 27.9~MB HiFi-GAN and 21.7~MB NOLA-\texttt{ConvTranspose1d} iSTFTNet specialists, and receptive-field window sizing paired with C++ \texttt{AudioBufferSourceNode} scheduling, Ninada achieves 3.09--3.22~dB 0--8~kHz MCD across 480 held-out studio utterances, 89--155~ms audible startup on desktop WebGPU, and 538--649~ms zero-underrun streaming on physical smartphone silicon.

{\small
\let\oldthebibliography\thebibliography
\renewcommand{\thebibliography}[1]{%
  \oldthebibliography{#1}%
  \setlength{\itemsep}{1pt}%
  \setlength{\parskip}{0pt}%
}
\bibliographystyle{plain}
\bibliography{references}
}

\end{document}